\documentclass[10pt]{iopart}
\usepackage{psfig}
\newcommand{\be}{\begin{equation}}
\newcommand{\ee}{\end{equation}}

\newcommand{\bea}{\begin{eqnarray}}
\newcommand{\eea}{\end{eqnarray}}
\renewcommand{\d}{{{\rm d}}}

\newcommand{\lton}{\mathrel{\lower.9ex
                  \hbox{$\stackrel{\displaystyle <}{\sim}$}}}

\begin{document}

\title[(Strange) Meson Interferometry at RHIC]
{(Strange) Meson Interferometry at RHIC}

\author{Sven Soff$\,^*$\footnote[3]{ssoff@lbl.gov}, Steffen~A.~Bass\dag, 
David~H.~Hardtke$\,^*$, and Sergey~Y.~Panitkin\ddag}

\address{$^*$ Nuclear Science Division 70-319, 
Lawrence Berkeley National Laboratory, 
One Cyclotron Road, Berkeley, CA94720, USA}

\address{\dag\ Department of Physics, Duke University, Durham,
NC27708, USA, and\\
RIKEN BNL Research Center, Brookhaven National Laboratory, 
Upton, NY11973, USA}

\address{\ddag\ Physics Department, Brookhaven National Laboratory, 
PO Box 5000, Upton, NY11973, USA}

\begin{abstract}
We make predictions for the kaon interferometry measurements in Au+Au 
collisions at the Relativistic Heavy Ion Collider (RHIC).
A first order phase transition from a thermalized Quark-Gluon-Plasma (QGP)
to a gas of hadrons is assumed for the transport calculations. 
The fraction of kaons that are directly emitted from 
the phase boundary is considerably enhanced at large transverse momenta 
$K_T \sim 1\,$GeV/c. In this kinematic region, the sensitivity 
of the $R_{\rm out}/R_{\rm side}$ ratio to the QGP-properties 
is enlarged. The results of the 1-dimensional correlation analysis are 
presented. The extracted interferometry radii, depending on $K_T$,  
are not unusually large and are strongly affected by finite 
momentum resolution effects. 
\end{abstract}
\vspace{-0.5cm}
\section{Introduction}
We discuss predictions for
the kaon interferometry measurements in Au+Au collisions at the
Relativistic Heavy Ion Collider (RHIC) that accelerates the nuclei up to 
nucleon-nucleon center-of-mass energies of $\sqrt{s}_{NN}=200\,$GeV.
Correlations of identical particle pairs, sometimes also called
HBT interferometry, provide important information on 
the space-time extension of the particle emitting source as produced 
for example in ultrarelativistic heavy ion
collisions \cite{Kopylov:1972qw}.
In this case, QCD lattice calculations have predicted
a transition from quark-gluon matter to hadronic matter at high
temperatures.
For a first-order phase transition, large hadronization times
have been expected due to the associated large latent heat as compared to
a purely hadronic scenario \cite{pratt86,schlei,dirk1}.
Entropy has to be conserved while the number of degrees of freedom
is reduced throughout the phase transition.
Thus, one has expected a considerable jump in the magnitude of
the interferometry radius parameters and the emission duration
once the energy density is large enough to produce 
quark-gluon matter.
Two alternative scenarios of the space-time evolution, 
with and without a phase transition, are
illustrated in Fig.~1 in the $z$-$t$-diagram.
After the collision of the two nuclei, each  with nucleon number $A$,
the system is formed at some eigen-time $\tau$ (indicated by the hyperbola)
and the initial expansion proceeds either in a hadronic state (left-hand side)
or in a state dominated by partonic degrees of freedom,
for example a quark-gluon plasma (QGP) (right-hand side).
In the latter case, the formation of a mixed phase, leads to
large hadronization times and thus to rather long emission durations.
The freeze-out is defined as the decoupling of
the particles, i.e., the space-time coordinates of their last
(strong) interactions.
Summarizing this idea illustrated in Fig.~1, the interferometry of 
identical particle pairs and in particular the
excitation function of the interferomtry parameters have been considered
as an ideal tool to detect the existence and the properties of a transition
from a thermalized quark-gluon plasma to hadrons.
Quantities of great interest are for example the critical 
temperature $T_c$ or the latent heat of the phase transition.
The prolonged emission duration should lead in particular to
an increase of the effective source size in the {\it outward} direction, 
i.e., parallel to the transverse pair velocity. 
One also expects correlation lengths depending on 
the specific entropy of the collision.
For recent reviews see for example~\cite{reviews,wiedemannrep}.

%%%%%%%%%%%%%%%%%%%%%%%%%%%%%%%%%
\parbox{7.8cm}{
\psfig{figure=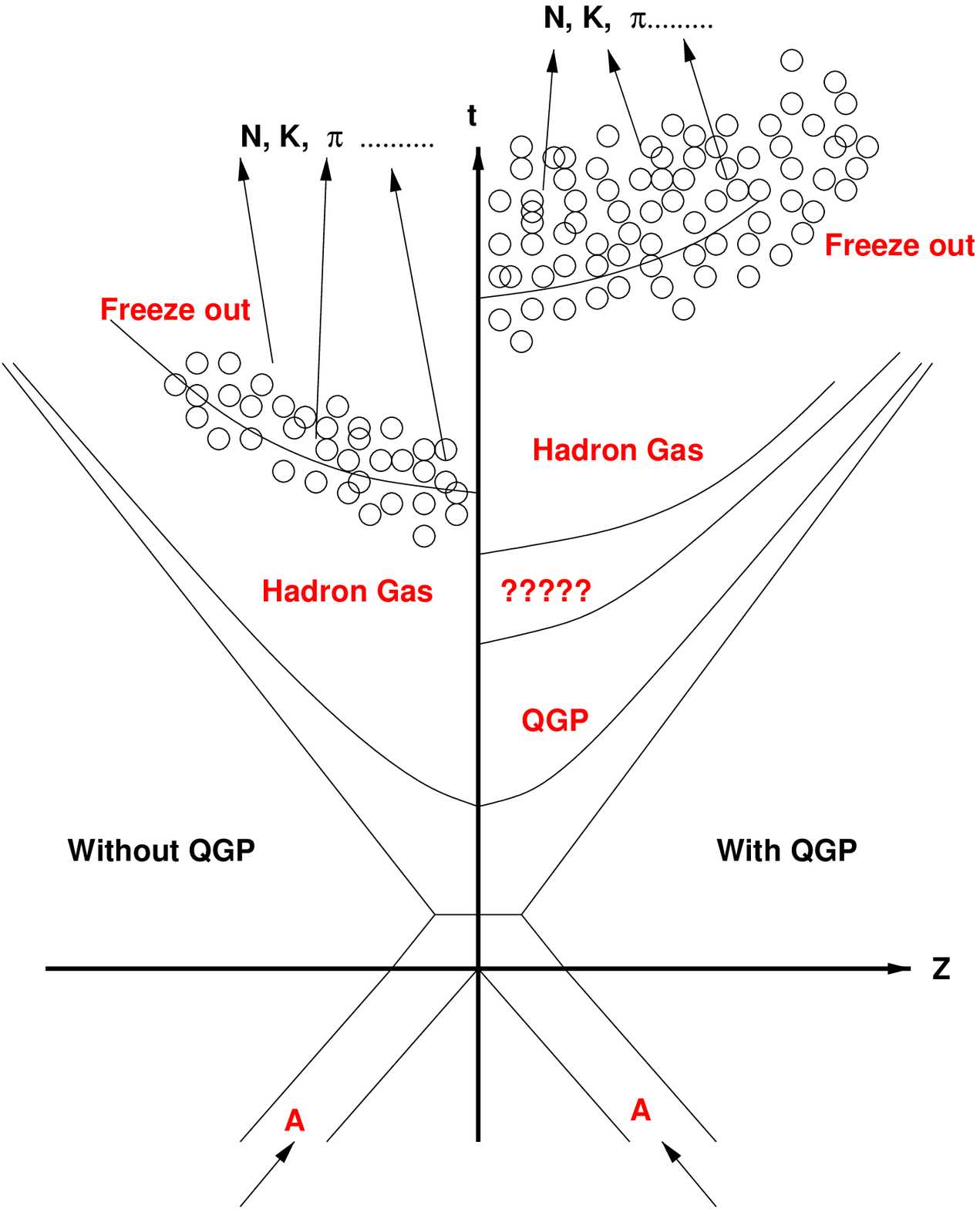,width=7.cm}
}
\parbox{5cm}{{\small Figure 1:
Illustration
of the space-time evolution in the
$z$-$t$-diagram with (right) and without (left) a phase
transition.
Proceeding through a first-order phase transition with a
large latent heat should
lead to large hadronization times, thus yielding eventually  
large interferometry radius parameters and emission duration.
}
}
%%%%%%%%%%%%%%%%%%%%%%%%%%%%%%%%

\section{Space-time evolution and kaon interferometry}
Here, we discuss calculations based on a relativistic 
two-phase dynamical transport
model that describes the early quark-gluon plasma phase
by hydrodynamics and the later stages, after hadronization from
the phase boundary of the mixed phase, by microscopic transport of the
hadrons. 
For the initial dense (hydrodynamical) phase
of a QGP a bag model
equation of state exhibiting a first order phase
transition is employed~\cite{dirk1,gersdorff}.
Hence, a phase
transition in local equilibrium that proceeds through the formation of a mixed
phase, is considered. 
A crossover \cite{dirk1,Zschiesche:2001dx} or a rapid out-of-equilibrium 
phase transition similar to spinodal decomposition~\cite{spino} 
may yield smaller radii and emission times.  
In the hadronic
phase, resonance (de)excitations and binary collisions are modeled based
on cross sections and resonance properties as measured in 
vacuum~\cite{bas98,soff00}.
A detailed description of this relativistic hybrid transport model and 
its predictions can be found elsewhere~\cite{DumRi,hu_main,kaonlett}.  
The particular differences and advantages compared to {\it pure} 
hydrodynamical calculations are for example due to the explicit calculation of 
the freeze-out phase-space distributions. Moreover,  
the system evolution in the later dilute stages is based 
on cross sections and resonance (de)excitations are possible. 
The assumption of {\it ideal} hydrodynamics might be not valid anymore 
in these later stages close to a freeze-out 
criterion like a fixed temperature $T_{\rm fr}$. 
Another important example will be shown below (Fig.~5), demonstrating 
the possibility of direct emission from the early phase in this 
transport model used here.  
The mean freeze-out times are strongly depending on the 
transverse momentum of the particles as illustrated in Fig.~2 for kaons 
and anti-kaons. Obviously, large transverse momentum kaons are 
correlated to early mean freeze-out times. 
The difference between kaons and anti-kaons is almost negligible, only 
at very small $p_t$ anti-kaons may be emitted a tiny instant later due 
a small finite net-baryon number in the system. (The 
K$^-$N cross section is considerably larger than the K$^+$N cross section.) 
Otherwise the K$\pi$ cross section, dominated by the K$^*$, yields 
almost identical freeze-out distributions for kaons and anti-kaons.
Strangeness distillation \cite{Greiner:1987tg} due to local large net-baryon 
numbers leading to emission asymetries that are observable 
in K$^+$K$^-$ correlations \cite{Soff:1997nb} is 
not considered here.
 
\parbox{7.3cm}{\psfig{figure=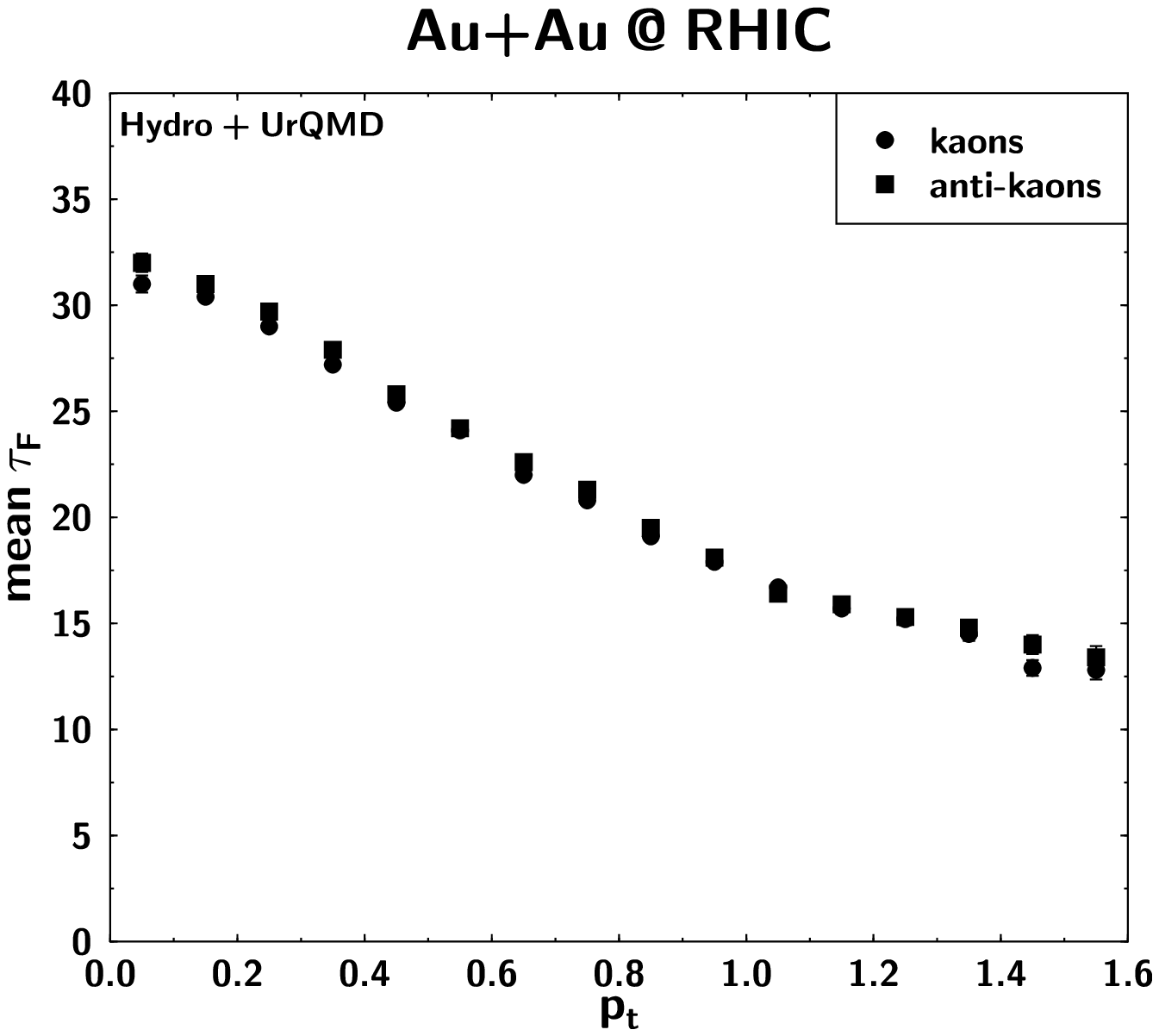,width=7.0cm}}
\parbox{5cm}{{\small Figure 2:
Mean freeze-out time $\tau_F$ as a function of transverse momentum $p_t$
for kaons and anti-kaons in Au+Au collisions at RHIC.
High $p_t$ kaons are correlated with early freeze-out times.}
}

The coordinate system for the correlation analysis is defined by  
the {\it long} axis ($z$) (parallel to
the beam axis), the {\it out} direction   
(parallel to the transverse momentum vector
${\bf K_T}=({\bf p_{T1}} + {\bf p_{T2}})/2$
of the pair), and the {\it side} direction (perpendicular
to both).
Due to the definition of the {\it out} and {\it side} direction,
$R_{\rm out}$ probes the spatial {\it and} temporal extension   
of the source while $R_{\rm side}$ only probes the spatial extension.
It has been suggested that the ratio $R_{\rm out}/
R_{\rm side}$ should increase strongly once the initial
entropy density $s_i$ becomes substantially larger than that of the hadronic
gas at $T_c$~\cite{dirk1}.
The  Gaussian radius parameters
can be obtained from a saddle-point integration over
the classical phase space distribution of the hadrons at freeze-out
(points of their last (strong) interaction) that is
identified with the Wigner density of the source,  
$S(x,K)$~\cite{Podgoretsky:1983xu,wiedemannrep,Gyulassy:1989yr,Pratt:1990zq}.
\begin{eqnarray}
\label{rs}
R_{\rm side}^2({\bf K_T})&=& \langle \tilde{y}^2 \rangle ({\bf K_T})\,,\\
R_{\rm out}^2({\bf K_T})&=& \langle (\tilde{x}-\beta_t \tilde{t})^2
\rangle ({\bf K_T}) = \langle\tilde{x}^2+\beta_t^2 \tilde{t}\,^2-2 
 \beta_t\tilde{x}\tilde{t}\rangle ({\bf K_T}) \,,\label{ro}\\
R_{\rm long}^2({\bf K_T})&=& \langle (\tilde{z}-\beta_l \tilde{t})^2
\rangle ({\bf K_T})\,,\label{rl}
\end{eqnarray}
with
\be
\tilde{x}^{\mu}({\bf K_T}) = x^{\mu} - \langle {x}^{\mu}\rangle({\bf K_T})
\ee
being the space-time coordinates relative
to the momentum dependent {\it effective source centers}.
The average in (\ref{rs})-(\ref{rl}) is taken over the   
emission function, i.e., 
\be
\langle f \rangle(K)= \frac{\int d^4x f(x) S(x,K)}{\int d^4x S(x,K)}
\ee
with $K=(E_K,{\bf K})$.
In the {\it osl} system ${\bf \beta}=(\beta_t,0,\beta_l)$, where
${\bf \beta}={\bf K}/E_K$ and $E_K=\sqrt{m^2+{\bf K}^2}$ 
(on-shell approximation).
For small $\tilde{x}$-$\tilde{t}$ correlations, i.e.\ in particular
at small $K_T$, $R_{\rm out}$ is increased relative
to $R_{\rm side}$ if the duration of emission $\Delta \tau = \surd
{\langle \tilde{t}\,^2 \rangle}$ is large~\cite{pratt86,schlei,dirk1}.
Strong (positive) $\tilde{x}$-$\tilde{t}$-correlations or large
spatial anisotropies in {\it out}- and {\it side}-direction
($\langle \tilde{y}^2\rangle > \langle \tilde{x}^2\rangle$) may,  
in principle, lead to $R_{\rm out} \le R_{\rm side}$. 
This is not seen within the model scenario discussed here. 

From eqs. (1) and (2) one can easily calculate the ratio 
$R_{\rm out}/R_{\rm side}$ for kaons. 
Kaons, compared to pions, are expected to be less contaminated 
by resonance decays~\cite{Gyulassy:1989yr,Sullivan:wb}. 
In addition the kaon phase space density is 
much smaller than the pion density~\cite{murray}. As a consequence, 
higher multiparticle correlation effects 
(that may be important for pions at RHIC energies~\cite{Lednicky:1999xz}) 
are possibly under much better control for kaons. 
Fig.~3 shows the kaon $R_{\rm out}/R_{\rm side}$ ratio for 
SPS (initial conditions $\tau_i=1\,$fm/c (thermalization eigen-time)
and $s/\rho_B=45$ (specific entropy density)) and 
RHIC ($\tau_i=0.6$, $s/\rho_B=200$)~\cite{kaonlett}. 
The bag parameter $B$ is varied from $380\,$MeV/fm$^3$ to 
$720\,$MeV/fm$^3$. This corresponds to a variation of the 
critical temperature from $T_c\approx 160\,$MeV to $T_c\approx
200\,$MeV. 

%%%%%%%%%%%%%%%%%%%%%%%%%%%%%%%%%%%%%%%%%
\parbox{7.3cm}{
\psfig{figure=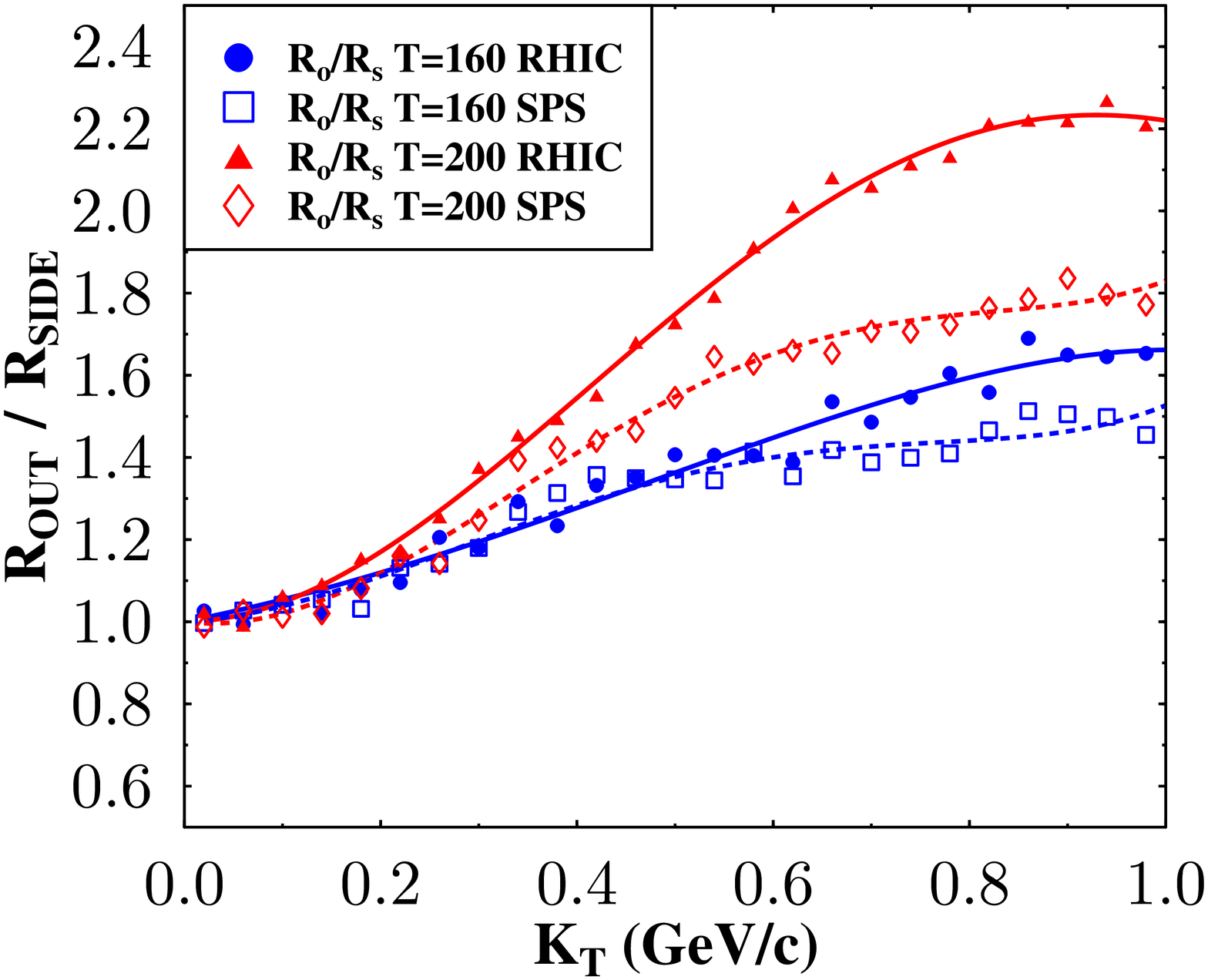,width=7.cm}    
}
\parbox{5cm}{{\small Figure 3: 
$R_{\rm out}/R_{\rm side}$ as obtained from eqs.\ (1) and (2) 
for kaons at RHIC (Au+Au, $\sqrt{s}_{NN}=200\,$GeV) (full symbols) and
at SPS (Pb+Pb, $\sqrt{s}_{NN}=17.4\,$GeV) (open symbols),
as a function of $K_T$ for critical temperatures  
$T_c\simeq 160\,$MeV  and $T_c\simeq 200\,$MeV, respectively.
}
}
%%%%%%%%%%%%%%%%%%%%%%%%%%%%%%%%%%%%%%%%

For small $K_T$, the ratio $R_{\rm out}/R_{\rm side}$ increases 
slowly compared to the faster rise for pions \cite{soffbassdumi}. 
This is due to the different masses yielding different flow velocities 
at the same $K_T$ (see eq.\ (2)). Apparently, 
the sensitivity to  the critical temperature  $T_c$ and the 
specific entropy density increases strongly with $K_T$. 
For higher $T_c$ the hadronization is faster but the 
subsequent hadronic rescattering phase lasts longer. 
It was shown that this dissipative hadronic phase dominates 
the radii for pions \cite{soffbassdumi}.
The ratio $R_{\rm out}/R_{\rm side}$ is always larger than 
unity and reaches values on the order of $\approx 1.5-2$ at large 
$K_T$. This value is considerably smaller than former  
expectations (e.g. \cite{dirk1,Bernard:1997bq}). 
On the other hand, first RHIC data for pion correlations 
show ratios that do not increase with $K_T$ and are even  
smaller than unity \cite{STARpreprint,Johnson:2001zi}. 
This completely new behaviour has  
not been seen at SPS energies (see e.g. \cite{blume01}). However, new 
data by the CERES collaboration show a similar trend \cite{appels02}. 
This observation would hint 
at a rather explosive scenario with
very short emission times, not compatible with a picture  
of a thermalized quark-gluon plasma hadronizing via a first-order
phase transition to an interacting hadron gas~\cite{soffbassdumi}.
Rather a shell-like emission as illustrated in Fig.~4 would be prefered.
Thus, the further study of two-particle interferometry will provide
extremly important information e.g.\ on the hadronization process or
the question of thermalization in ultrarelativistic heavy ion collisions.
Also, the comparison of pion and kaon interferometry data may 
further clarify the situation. 

%%%%%%%%%%%%%%%%%%%%%%%%%
\parbox{7.4cm}{
\psfig{figure=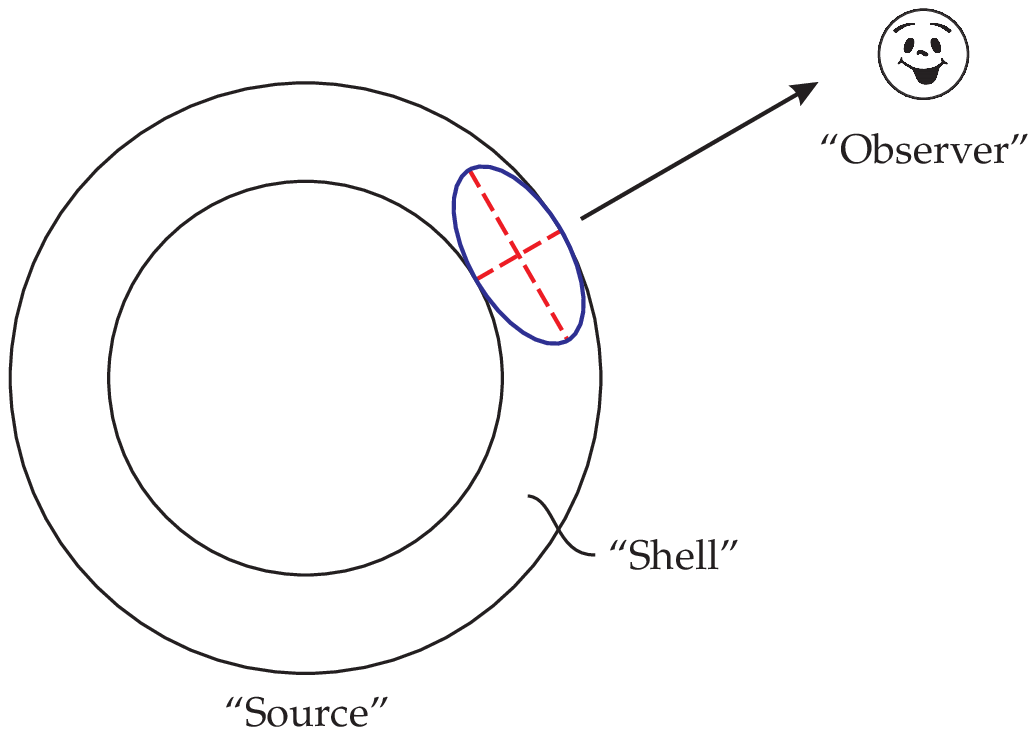,width=7.6cm}
}
\parbox{5.2cm}{{\small Figure 4: 
Illustration of a shell-like emission.
The surface emission geometry corresponds to small values
of the ratio $R_{\rm out}/R_{\rm side}$, indicated by the two dashed lines
in the emission volume element relevant for an observer.
The dashed line in the direction to the observer corresponds to the
{\it out} direction, the orthogonal line is the {\it side} direction.
}
}
%%%%%%%%%%%%%%%%%%%%%%%%%
 
Kaons are less distorted by
long-lived resonances and escape the opaque hadronic phase easier 
(compared to pions). 
About $30\%$ of the kaons at $K_T\sim 1\,$GeV/c are directly emitted
from the phase-boundary (for $T_c\approx160\,$MeV) as can be seen 
in Fig.~5. Complementary, we have seen already that large $K_T$ kaons and 
their $R_{\rm out}/R_{\rm side}$ ratio
exhibit a strong sensitivity on the QCD equation of state. 
The fraction of resonance decays (K$^*$'s,$\ldots$) is still quite large and 
decreases with $K_T$ from about 70$\%$ to 50$\%$. 
For the higher $T_c$ ($\simeq 200\,$MeV), hadronization is earlier and 
the hadronic phase lasts longer, such that the system is more opaque   
for direct emission than in the lower $T_c$ case.

%%%%%%%%%%%%%%%%%%%%%%%%%%%%%%%%%%%%
\parbox{12cm}{
\psfig{figure=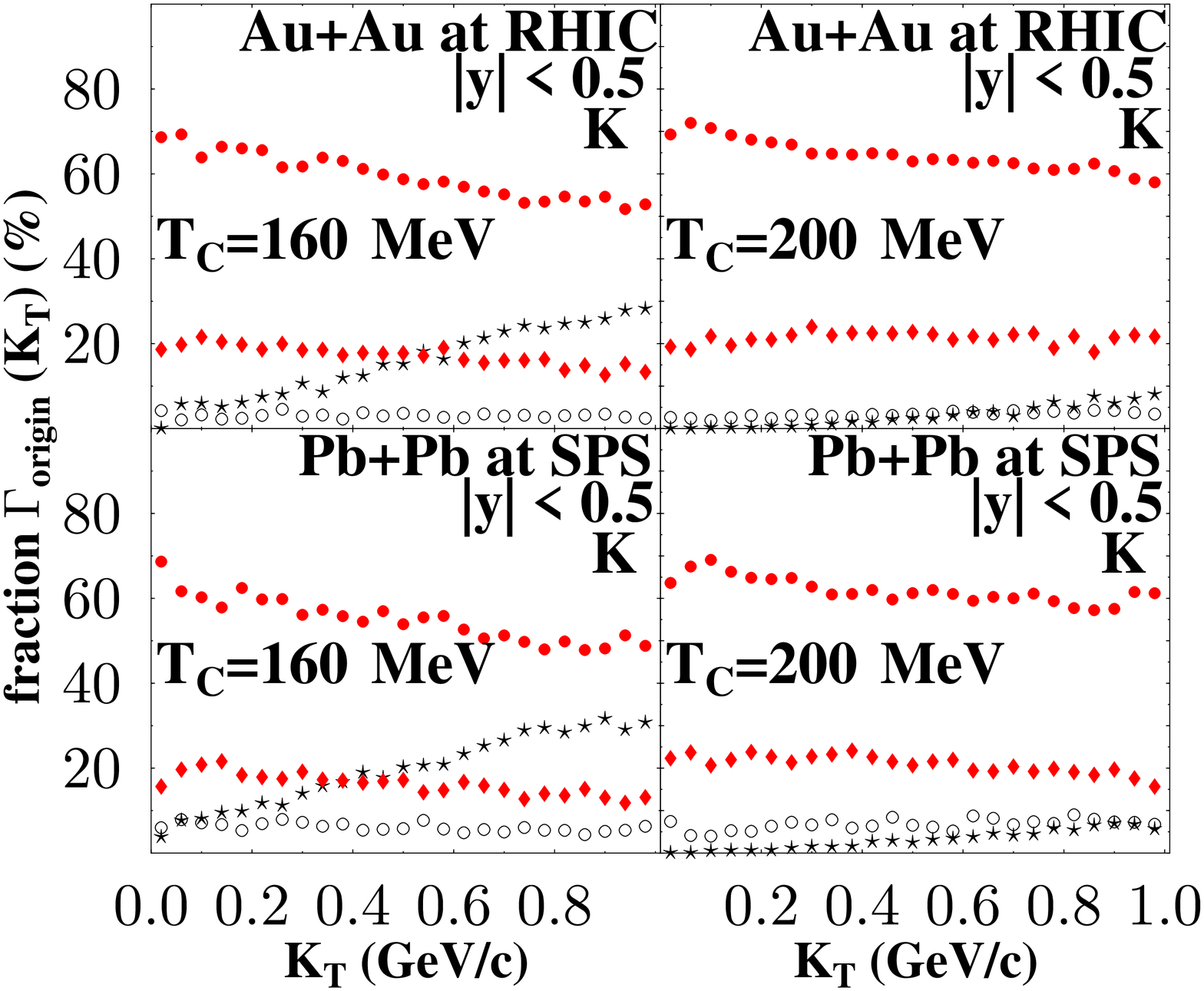,width=9.5cm}
}

\parbox{12cm}{{\small Figure 5: 
Fraction of kaons $\Gamma_{\rm origin}$ that origin from a particular
reaction channel prior to freeze-out. These are resonance
decays (full circles),  direct emission from the phase boundary (stars),
elastic meson-meson (diamonds), or elastic meson-baryon (open circles)  
collisions. The upper and lower diagrams are for RHIC 
and SPS initial conditions for
$T_c\simeq 160\,$MeV (left) and $T_c\simeq 200\,$MeV (right), respectively.
}
}
%%%%%%%%%%%%%%%%%%%%%%%%%%%%%%%%%%%%

We now calculate the one-dimensional kaon kaon correlation functions 
$C_2(Q_{\rm inv})$ for various $K_T$-bins and  
determine the corresponding fit parameters. 
The corresponding results for the three-dimensional analysis in the 
{\it out-side-long} coordinate system are presented in \cite{kaonlett}. 
The parameters $R_{\rm inv}$ and $\lambda_1$ of the 
correlation functions are obtained by fitting  a Gaussian as 
\begin{equation}
C_2(Q_{\rm inv})=1+\lambda_1 \exp(-Q_{\rm inv}^2 R_{\rm inv}^2)\,.
\end{equation}
The correlation functions themselves are calculated from
the phase space distributions of kaons at freeze-out
using the {\it correlation after burner} by Pratt~\cite{pratt86,Pratt:1990zq}.
The correlation functions can be calculated as  
\begin{equation}
C_2({\bf p_1},{\bf p_2}) \simeq 1 + \frac
{\int\d^4x S(x,{\bf K}) \int\d^4y S(y,{\bf K}) \exp\left(
2ik\cdot(x-y)\right)}
{|\int\d^4x S(x,{\bf K})|^2}\,.
\end{equation}
with $2{\bf K}={\bf p_1}+{\bf p_2}$, $2{\bf k}={\bf p_1}-{\bf p_2}$, and
$2k^0 = E_{p_1}-E_{p_2}$.

%%%%%%%%%%%%%%%%%%%%%%%%%%%%%%%%%%%%%
\parbox{12cm}{
\psfig{figure=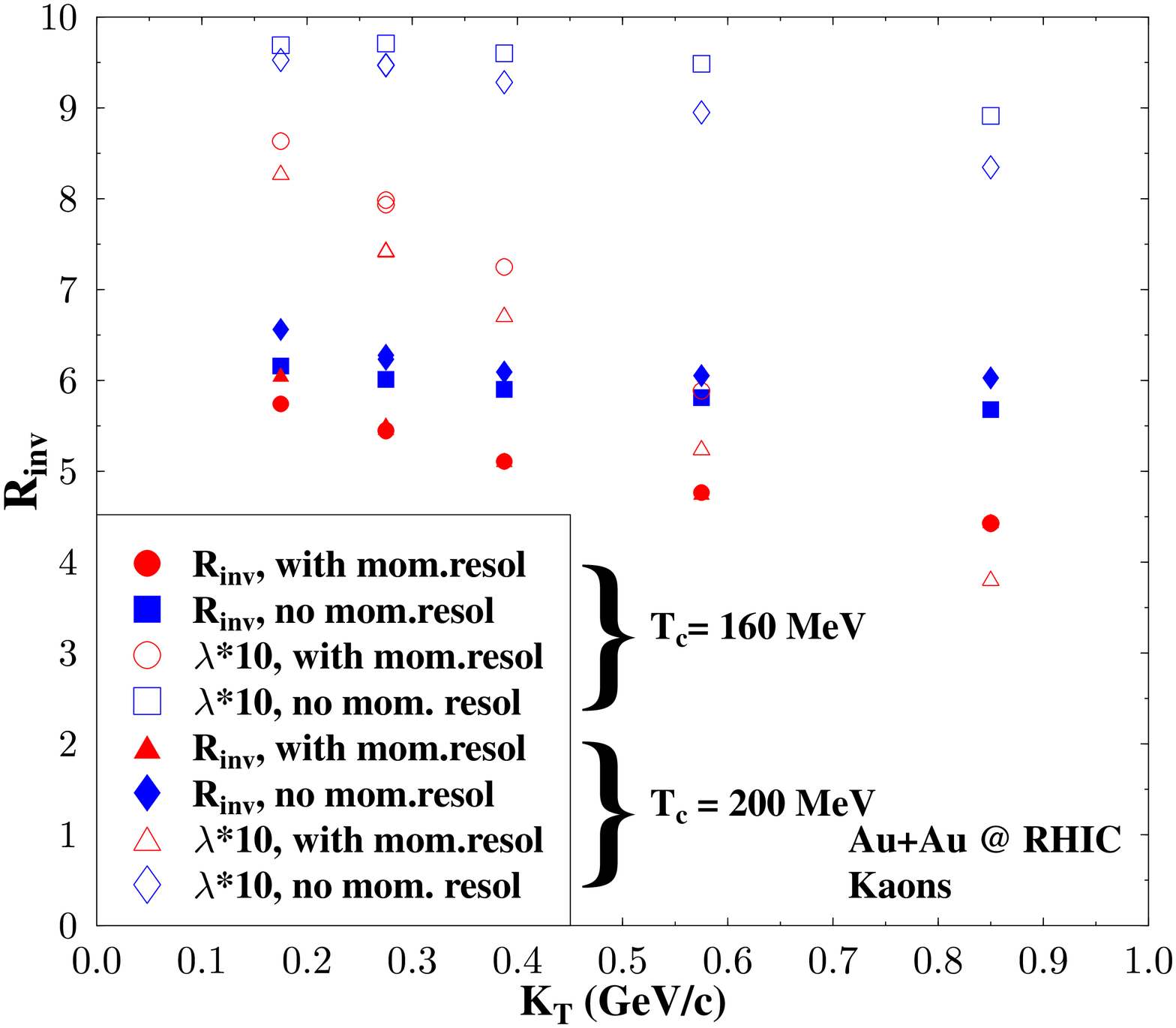,width=10.cm}
}

\parbox{12cm}{{\small Figure 6: 
Parameters $R_{\rm inv}$ and $\lambda_1$ (as obtained from a fit 
according to eq. (6))  
as a function of $K_T$ for kaons in Au+Au collisions at RHIC for critical
temperatures $T_c\simeq 160\,$MeV (circles, squares) and
$T_c\simeq 200\,$MeV (triangles, diamonds), respectively.
Calculations with and without taking momentum resolution effects into account
are shown.
}
}
%%%%%%%%%%%%%%%%%%%%%%%%%%%%%%%%%%

The results for RHIC are shown in Fig.~6. The $\lambda_1$ intercept
parameters are multiplied by a factor 10. 
The correlation functions have been calculated for 
$T_c\approx160\,$MeV (circles and squares) and $T_c\approx200\,$MeV 
(triangles and diamonds). Moreover, calculations with and without 
taking finite momentum resolution ({\it f.m.r.)} effects have been 
performed. 
In the first case, 
the {\it true} particle momentum $p$
obtains an additional random component. This random component is
assumed to be Gaussian with a width $\delta p$.
A {\it f.m.r.\ }of $\approx 2\%$ of the center of
each $K_T$ bin is considered, 
a value assumed for the STAR detector \cite{STARpreprint}.
The relative momenta of pairs are then calculated from these modified momenta.
However, the correlator is calculated with the {\it true} relative
momentum. 
$R_{\rm inv}$ is on the order of $5.5-6.5\,$fm without and 
$4.5-6\,$fm with taking  {\it f.m.r.} into account. 
Hence, $R_{\rm inv}$ is reduced by the {\it f.m.r.}. 
This reduction grows with $K_T$. 
For the one-dimensional fit parameters, there is hardly a sensitivity 
to the critical temperature. 
However, the known trend 
(larger radii for higher $T_c$ due to a prolonged hadronic phase) 
is also indicated in these results. 
The sensitivity to the specific entropy density, another initial condition, 
is known to be rather weak \cite{soffbassdumi}, even for a  change from SPS 
to RHIC initial conditions, due to the dominance of the late soft hadronic 
interactions. Hence, the differences in the correlation parameters 
between, for example, the RHIC 130$\,$GeV and 200$\,$GeV energies 
are minute; (the number of charged particles at midrapidity increases only 
by about $15\%$). 
Preliminary analysis of STAR KK data (not yet corrected for {\it f.m.r.}) at 
$\sqrt{s}_{NN}=130\,$GeV gives a value of $R_{\rm inv}=4.5\pm 0.3\,$fm 
(at $150\,$MeV$<K_T<400\,$MeV and rather central (11$\%$) 
collisions) ($\lambda_1=0.92\pm0.13$)~\cite{panitkinhere}, which is 
almost compatible 
to the calculations that take {\it f.m.r.} into account.  
The $\lambda_1$ intercept parameters are almost constant without 
{\it f.m.r.} effects but strongly decrease when taking them into
account. 
For illustration, let us consider
particle pairs with a small {\it true} relative momentum $q$
that will be redistributed by the random momentum smearing to,
on the average, larger relative momenta $\tilde{q}$. Thus, the
area around $q\approx 0$ is depleted from these {\it true} low momentum
pairs that carry the full correlation strength. Complementary,
some pairs with a larger {\it true} $q$ (and thus with a weaker
correlation strength) are redistributed to
the small $\tilde{q}$ area. As a consequence, the {\it correlation
strength is transported to larger $q$ values}.
Thus, the correlation function gets broader and the radius parameters become
smaller.
The strong decrease of $\lambda_1$ with $K_T$ is due to the absolute
larger momentum smearing at high $K_T$. {\it True} correlated pairs at
low $q$ are transported more likely  to larger $\tilde{q}$ values.
This reduced correlation strength in the first $q$ bins causes the small
$\lambda_1$ values at larger $K_T$.
Recent experimental data (corrected for {\it f.m.r.}!) 
for central Pb+Pb collisions at
$\sqrt{s}=17.4\,$GeV yield
$R_{\rm inv}=6.2\pm 0.4\pm 0.9\,$fm at low transverse momenta
($\langle p_t \rangle = 250\,$MeV) and
$R_{\rm inv}=3.3 \pm 0.6\,$fm at high transverse momenta
($\langle p_t \rangle = 910\,$MeV) \cite{na44prlnew}.
The corresponding $\lambda_1$ parameters are reported as
$\lambda_1=0.84\pm0.11\pm0.19$ and $\lambda_1=0.47\pm0.12$ for
the low and high $p_t$ set, respectively.
\section{Summary}
\vspace{-.2cm}
The calculation of kaon correlation parameters for Au+Au collisions 
at RHIC energies, assuming a first-order phase
transition from a thermalized QGP to a gas of hadrons shows  
\begin{itemize}
\item 
an increasing $R_{\rm out}/R_{\rm side}$ ratio with $K_T$ (larger than 
unity),
\item
an increasing sensitivity of the $R_{\rm out}/R_{\rm side}$ ratio with 
$K_T$,
\item
a strong ($30\%$) direct emission component from the phase boundary at 
large $K_T\sim 1\,$GeV/c,
\item
no unusually large radius parameters $R_{\rm inv}$ that are reduced 
due to finite momentum resolution effects,
\item 
$\lambda_1$ intercept parameters that get strongly 
reduced due to finite momentum resolution, 
\item 
the reduction of $R_{\rm inv}$ and $\lambda_1$ grows with $K_T$.
\end{itemize}
% \vspace*{-0.5cm}
The kaon interferometry measurements at RHIC (at high $K_T$) 
will be, in combination with the pion data, 
an utmost important and valuable  
probe of the space-time dynamics (close to the phase
boundary) and to the properties of the phase transition.  
\vspace*{-0.4cm}
\ack
\vspace*{-0.3cm}
This work is supported by the Alexander von Humboldt-Foundation through 
a Feodor Lynen Fellowship (SS) and the 
Gesellschaft f\"ur Schwerionenforschung (Darmstadt).  
We thank A.\ Dumitru for many helpful 
comments, the 
UrQMD collaboration for permission to use the UrQMD transport model
and S.\ Pratt for providing the correlation program CRAB.
SS acknowledges support from 
DOE Grant No.\ DE-AC03-76SF00098 and 
SAB from DOE Grant No.\ DE-FG02-96ER40945
and DE-AC02-98CH10886.
%This work was supported by the Director, Office of Energy Research,
%Office of High Energy and Nuclear Physics, Division of Nuclear
%Physics, of the U.S. Department of Energy under
%Contract No.\ DE-AC03-76SF00098.

\end{document}